\documentclass[twoside,twocolumn,9pt]{article}
\usepackage{extsizes}
\usepackage[super,sort&compress,comma]{natbib} 
\usepackage[version=3]{mhchem}
\usepackage[left=1.5cm, right=1.5cm, top=1.785cm, bottom=2.0cm]{geometry}
\usepackage{balance}
\usepackage{mathptmx}
\usepackage{sectsty}
\usepackage{graphicx} 
\usepackage{lastpage}
\usepackage[format=plain,justification=justified,singlelinecheck=false,font={stretch=1.125,small,sf},labelfont=bf,labelsep=space]{caption}
\usepackage{float}
\usepackage{fancyhdr}
\usepackage{fnpos}
\usepackage[english]{babel}
\addto{\captionsenglish}{%
  \renewcommand{\refname}{Notes and references}
}
\usepackage{array}
\usepackage{droidsans}
\usepackage{charter}
\usepackage[T1]{fontenc}
\usepackage[usenames,dvipsnames]{xcolor}
\usepackage{setspace}
\usepackage[compact]{titlesec}
\usepackage{hyperref}
\usepackage{stfloats}
\usepackage{amssymb}
\usepackage{epstopdf}%This line makes .eps figures into .pdf - please comment out if not required.
\usepackage{wrapfig}

\makeatletter

\newcommand{\Rmnum}[1]{\expandafter\@slowromancap\romannumeral #1@}
\makeatother

\definecolor{cream}{RGB}{222,217,201}

\begin{document}

\pagestyle{fancy}
\thispagestyle{plain}
\fancypagestyle{plain}{
%%%HEADER%%%
\renewcommand{\headrulewidth}{0pt}
}
%%%END OF HEADER%%%

%%%PAGE SETUP - Please do not change any commands within this section%%%
\makeFNbottom
\makeatletter
\renewcommand\LARGE{\@setfontsize\LARGE{15pt}{17}}
\renewcommand\Large{\@setfontsize\Large{12pt}{14}}
\renewcommand\large{\@setfontsize\large{10pt}{12}}
\renewcommand\footnotesize{\@setfontsize\footnotesize{7pt}{10}}
\makeatother

\renewcommand{\thefootnote}{\fnsymbol{footnote}}
\renewcommand\footnoterule{\vspace*{1pt}% 
\color{cream}\hrule width 3.5in height 0.4pt \color{black}\vspace*{5pt}} 
\setcounter{secnumdepth}{5}

\makeatletter 
\renewcommand\@biblabel[1]{#1}            
\renewcommand\@makefntext[1]% 
{\noindent\makebox[0pt][r]{\@thefnmark\,}#1}
\makeatother 
\renewcommand{\figurename}{\small{Fig.}~}
\sectionfont{\sffamily\Large}
\subsectionfont{\normalsize}
\subsubsectionfont{\bf}
\setstretch{1.125} %In particular, please do not alter this line.
\setlength{\skip\footins}{0.8cm}
\setlength{\footnotesep}{0.25cm}
\setlength{\jot}{10pt}
\titlespacing*{\section}{0pt}{4pt}{4pt}
\titlespacing*{\subsection}{0pt}{15pt}{1pt}
%%%END OF PAGE SETUP%%%

%%%FOOTER%%%
\fancyfoot{}
\fancyfoot[LO,RE]{\vspace{-7.1pt}\includegraphics[height=9pt]{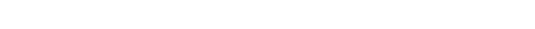}}
\fancyfoot[CO]{\vspace{-7.1pt}\hspace{13.2cm}\includegraphics{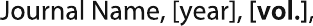}}
\fancyfoot[CE]{\vspace{-7.2pt}\hspace{-14.2cm}\includegraphics{head_foot/RF}}
\fancyfoot[RO]{\footnotesize{\sffamily{1--\pageref{LastPage} ~\textbar  \hspace{2pt}\thepage}}}
\fancyfoot[LE]{\footnotesize{\sffamily{\thepage~\textbar\hspace{3.45cm} 1--\pageref{LastPage}}}}
\fancyhead{}
\renewcommand{\headrulewidth}{0pt} 
\renewcommand{\footrulewidth}{0pt}
\setlength{\arrayrulewidth}{1pt}
\setlength{\columnsep}{6.5mm}
\setlength\bibsep{1pt}
%%%END OF FOOTER%%%

%%%FIGURE SETUP - please do not change any commands within this section%%%
\makeatletter 
\newlength{\figrulesep} 
\setlength{\figrulesep}{0.5\textfloatsep} 

\newcommand{\topfigrule}{\vspace*{-1pt}% 
\noindent{\color{cream}\rule[-\figrulesep]{\columnwidth}{1.5pt}} }

\newcommand{\botfigrule}{\vspace*{-2pt}% 
\noindent{\color{cream}\rule[\figrulesep]{\columnwidth}{1.5pt}} }

\newcommand{\dblfigrule}{\vspace*{-1pt}% 
\noindent{\color{cream}\rule[-\figrulesep]{\textwidth}{1.5pt}} }

\makeatother
%%%END OF FIGURE SETUP%%%

%%%TITLE, AUTHORS AND ABSTRACT%%%
\twocolumn[
  \begin{@twocolumnfalse}
{\includegraphics[height=30pt]{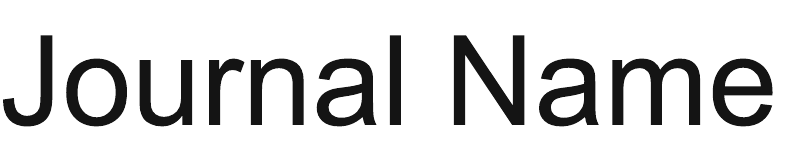}\hfill\raisebox{0pt}[0pt][0pt]{\includegraphics[height=55pt]{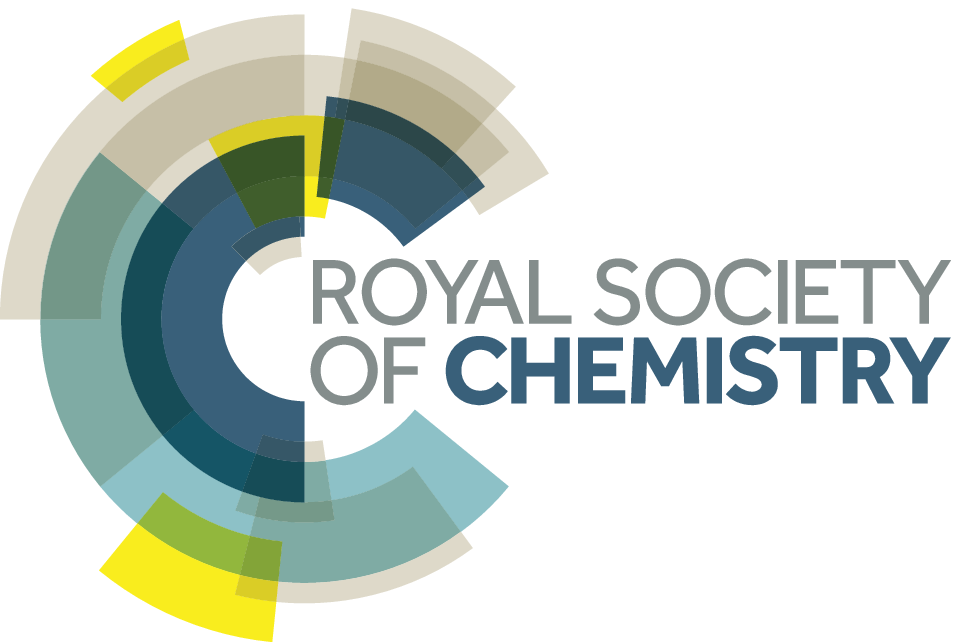}}\\[1ex]
\includegraphics[width=18.5cm]{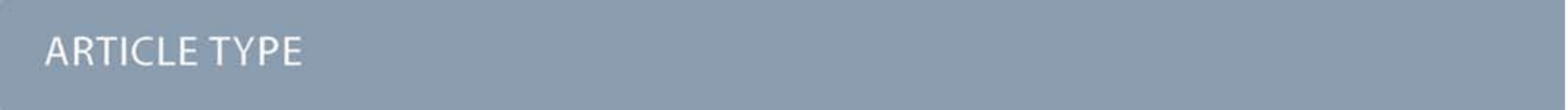}}\par
\vspace{1em}
\sffamily
\begin{tabular}{m{4.5cm} p{13.5cm} }

\includegraphics{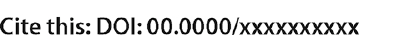} & \noindent\LARGE{\textbf{Locally Spontaneous Dynamic Oxygen Migration on Biphenylene: A DFT Study$^\dag$}} \\%Article title goes here instead of the text "This is the title"
\vspace{0.3cm} & \vspace{0.3cm} \\

 & \noindent\large{Boyi Situ,\textit{$^{a\ddag}$} Zihan Yan,\textit{$^{a\ddag}$} Rubin Huo,\textit{$^{a}$} Kongbo Wang,\textit{$^{a}$} Liang Chen,\textit{$^{b}$} Zhe Zhang,\textit{$^{a}$} Liang Zhao,$^{\ast}$\textit{$^{a}$} and Yusong Tu$^{\ast}$\textit{$^{a}$}} \\%Author names go here instead of "Full name", etc.

\includegraphics{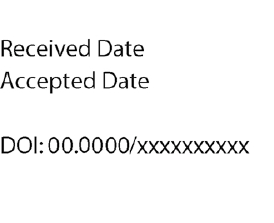} & \noindent\normalsize{The dynamic oxygen migration at the interface of carbon allotropes dominated by the periodic hexagonal rings, including graphene and carbon nanotube, has opened up a new avenue to realize dynamic covalent materials. However, for the carbon materials with hybrid carbon rings, such as biphenylene, whether the dynamic oxygen migration at its interface can still be found remains unknown. Using both density functional theory calculations and machine-learning-based molecular dynamics (MLMD) simulations, we found that the oxygen migration departing away from the four-membered carbon ($C_4$) ring is hindered, and the oxygen atom prefers to spontaneously migrate toward/around the $C_4$ ring. This locally spontaneous dynamic oxygen migration on the biphenylene is attributed to the high barrier of about 1.5 eV for the former process and relatively low barrier of about 0.3 eV for the latter one, originating from the enhanced activity of C-O bond near/around the $C_4$ ring due to the hybrid carbon rings structure. Moreover, the locally spontaneous dynamic oxygen migration is further confirmed by MLMD simulations. This work sheds light on the potential of biphenylene as a catalyst for spatial-controlled energy conversion and provides the guidance for realizing the dynamic covalent interface at other carbon-based or two-dimensional materials.} \\%The abstrast goes here instead of the text "The abstract should be..."

\end{tabular}

 \end{@twocolumnfalse} \vspace{0.6cm}

  ]
%%%END OF TITLE, AUTHORS AND ABSTRACT%%%

%%%FONT SETUP - please do not change any commands within this section
\renewcommand*\rmdefault{bch}\normalfont\upshape
\rmfamily
\section*{}
\vspace{-1cm}

%%%FOOTNOTES%%%

\footnotetext{\textit{$^{a}$~College of Physics Science and Technology \& Microelectronics Industry Research Institute, Yangzhou University, Jiangsu 225009, China. E-mail: zhaoliang@yzu.edu.cn; ystu@yzu.edu.cn}}
\footnotetext{\textit{$^{b}$~School of Physical Science and Technology, Ningbo University, Ningbo 315211, China.}}

%Please use \dag to cite the ESI in the main text of the article.
%If you article does not have ESI please remove the the \dag symbol from the title and the footnotetext below.
\footnotetext{\dag~Electronic Supplementary Information (ESI) available. }
%additional addresses can be cited as above using the lower-case letters, c, d, e... If all authors are from the same address, no letter is required

\footnotetext{\ddag~Boyi Situ and Zihan Yan contributed equally to this work.}

%%%END OF FOOTNOTES%%%

%%%MAIN TEXT%%%%
\section{Introduction}

Understanding the oxygen-related dynamic behavior at the interface of two-dimensional (2D) materials is of fundamental importance for the advancements of dynamic covalent materials \cite{chakma2019dynamic,roy2015dynamers,zou2017dynamic,tu2020water,yan2022remarkably} and the innovations of technological applications in sensors, \cite{lee2016biosensors,li2015graphene} coatings, \cite{zhang2016substrate,wang2019dhq,prasai2012graphene} and batteries. \cite{yan2012first,jung2019rapid,tian2017fast} As a widely used carbon-based 2D material, graphene oxide (GO) is abundant with oxygen-containing functional groups such as epoxy and hydroxyl, on its basal plane. \cite{zhu2010graphene,chen2012graphene,yang2014high,huang2020physical} Conventionally, these groups can seldom migrate or move along the GO interface due to the high oxygen migration barrier. \cite{zhang2016substrate,alihosseini2023solvent} Unexpectedly, we have found that with the mediation of water molecules, the oxygen functional groups can spontaneously migrate along the GO interface at room temperature via the C-O bond breaking/reforming and the proton transfer between a dangling oxygen and a neighboring hydroxyl group, which is attributed to the relatively low oxygen migration barrier. This dynamic oxygen migration on GO interface leads to the structural adaptivity to biomolecule adsorption \cite{tu2020water} and facilitates the ultrafast response to environmental humidity.\cite{zeng2023ultrafast} Even without the water, the spontaneously dynamic oxygen migration can still be observed at the interface of carbon nanotube \cite{zhu2021unexpected} or be remarkably enhanced on the copper-substrate supported GO.\cite{yan2022remarkably} These previous works all concern the carbon-based materials with periodic hexagonal rings, however, whether the dynamic oxygen migration on those with hybrid carbon rings can be accessed remains unknown.

Beyond graphene dominated by the periodic hexagonal rings, several carbon allotropes composed of hybrid carbon rings have been recently predicted by first-principles calculations or synthesized experimentally.\cite{fan2021biphenylene,liu2012structural,liu2017graphene,zhang2015penta} Biphenylene, as a novel 2D carbon material consisting of four-($C_4$), six-($C_6$) and eight-membered ($C_8$) carbon rings, shows great potential in applications such as catalysis,\cite{luo2021first,takano2019recent} batteries\cite{al2022two} and sensors,\cite{su2022theoretical,hosseini2020theoretical,mahamiya2022remarkable} etc. The hybrid carbon rings at the biphenylene monolayer surface introduce non-equivalent carbon sites and improve the interfacial activity, such that the biphenylene monolayer can be easily oxidized and shows more active carbon sites for chemical reactions.\cite{su2022theoretical,liu2021two} It is thus expected that the oxygen groups on biphenylene monolayer could also show dynamic behaviors, similar to those on GO.

In this work, we explore the dynamic oxygen migration on the biphenylene monolayer using the combination of density functional theory (DFT) calculations and machine-learning-based molecular dynamics (MLMD) simulations. It is found that the oxygen migration departing away from the $C_4$ ring is hindered and the oxygen prefers to spontaneously migrate toward/around the $C_4$ ring, leading to the locally spontaneous dynamic oxygen migration on the biphenylene monolayer. To the best of our knowledge, this is the first report on the interfacial oxygen-related dynamic behavior on the biphenylene monolayer, which predicts that biphenylene monolayer is an excellent candidate for 2D dynamic covalent interface and lays the foundation for further applications, such as oxygen-related catalysis and (bio)sensors.

\section{Computational model and methods}

\subsection{Model of biphenylene monolayer} 

\begin{figure}[]
	\centering
	\includegraphics[width=0.44\textwidth]{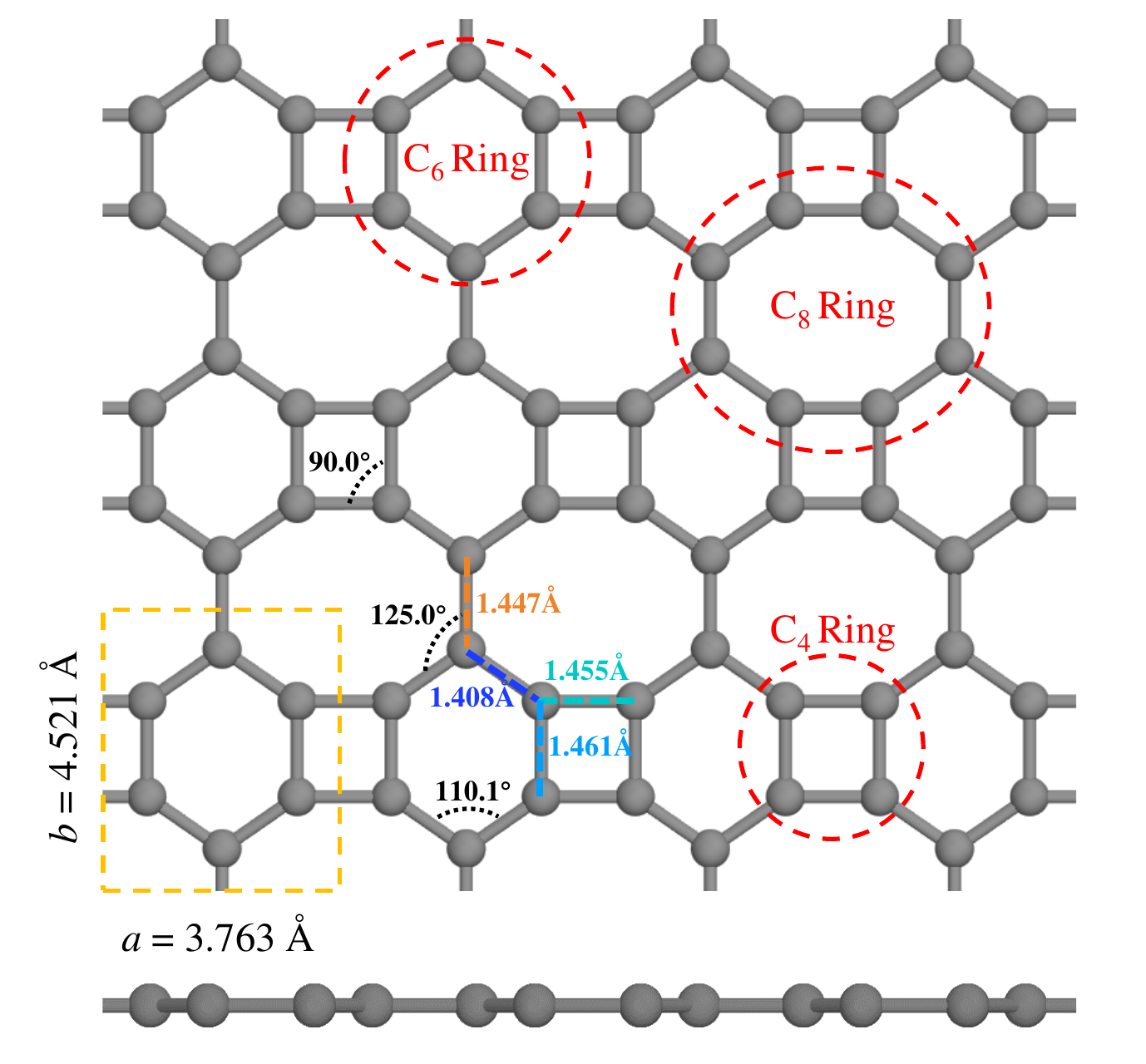}
	\caption{Top and side views of the optimized atomic structure of the biphenylene monolayer. The primitive unit cell is indicated by the yellow dashed box, the $C_4$, $C_6$ and $C_8$ rings are circled by red dashed rings. The C-C bonds between adjacent carbon rings are marked by dashed lines with different colors, and angles are marked by black dashed curves.}
	\label{figure1}
\end{figure}

\begin{figure*}[bp]
	\centering
	\includegraphics[width=0.75\textwidth]{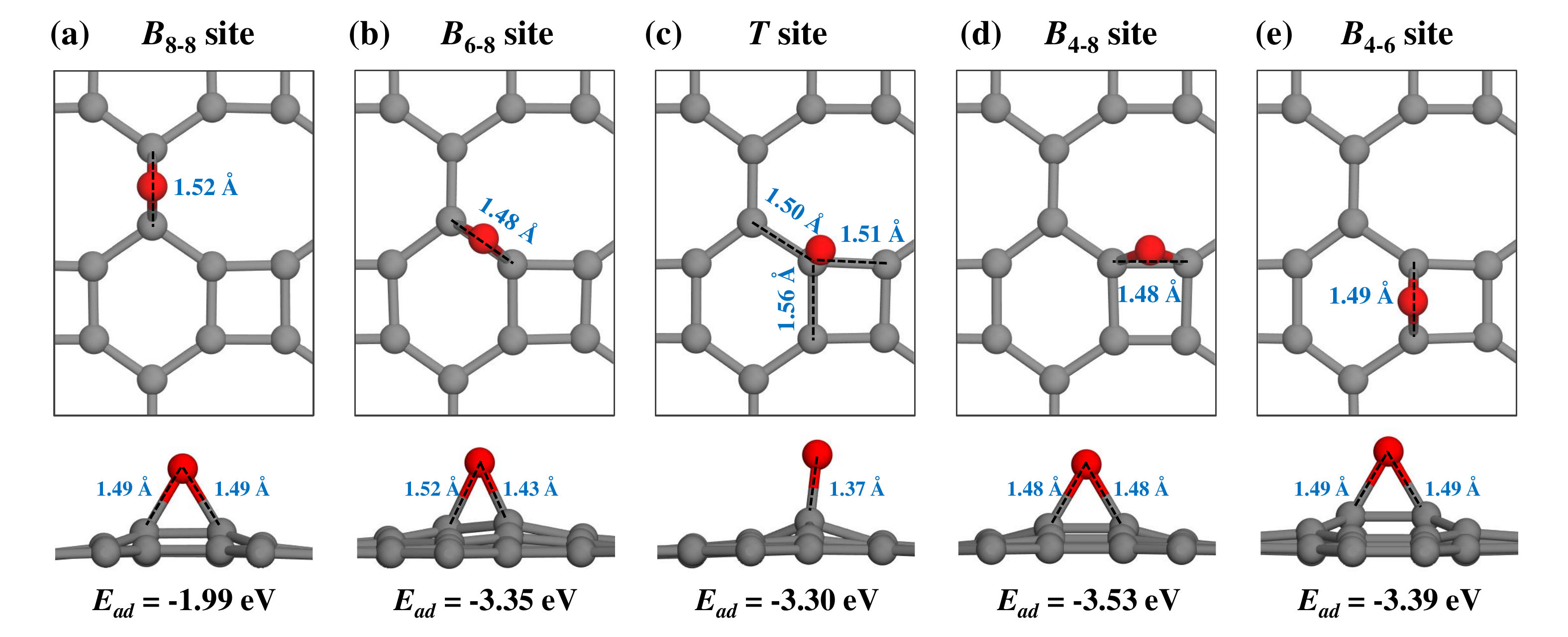}
	\caption{Top and side views of the optimized structures of an oxygen atom chemisorbed on five different non-equivalent carbon sites: (a) shared C-C bond between the adjacent two $C_8$ rings ($B_{8-8}$ site), (b) adjacent $C_6$-$C_8$ rings ($B_{6-8}$ site), (c) shared carbon atom of $C_4$-$C_6$-$C_8$ rings ($T$ site), (d) shared C-C bond between adjacent $C_4$-$C_8$ rings ($B_{4-8}$ site) and (e) adjacent $C_4$-$C_6$ rings ($B_{4-6}$ site). The blue numbers denote the bond lengths and the adsorption energy is denoted by Ead with respect to the isolated units.}
	\label{figure2}
\end{figure*}

Fig. \ref{figure1} shows the biphenylene monolayer including 72 carbon atoms within a $4\times3$ supercell. The minimal cell structure contains 6 carbon atoms, with optimized lattice constants of a = 3.763 \AA\ and b = 4.521 \AA. Unlike the primitive hexagonal structure of graphene with the C-C bond length of 1.426 \AA,\cite{dai2013diffusion} the biphenylene monolayer is composed of $C_4$, $C_6$ and $C_8$ rings, and shows different C-C bond lengths ranging from 1.408 \AA\ to 1.461 \AA, well matching with the reported results.\cite{mahamiya2022remarkable}

\subsection{Computational details}
All the calculations were performed at the first-principles levels within the framework of DFT, implemented in the Vienna \textit{Ab initio} Simulation Package (VASP).\cite{kresse1996efficient,kresse1994ab} Projector-augmented wave (PAW) method\cite{blochl1994projector} was used to describe the electron-ion interaction with the plane-wave energy cutoff of 500 eV. The exchange-correlation energy was described by the generalized gradient approximation (GGA) with the Perdew-Burke-Ernzerhof (PBE) exchange functional,\cite{perdew1996generalized} and Grimme’s DFT-D3 correction\cite{grimme2010consistent} was adopted to describe the nonlocal dispersive interactions. The convergence accuracy for energy was $10^{-5}$ eV within the spin-polarized calculations, and force on each atom was less than 0.01 eV/\AA. The Gamma-centered $3\times3\times1$, $9\times9\times1$ and $9\times9\times1$ \textit{k}-point mesh in the Brillouin zone were used for geometry optimization, self-consistent field calculation and density of states analysis, respectively. The vacuum layer in the \textit{z}-axis direction was 20\AA\ to avoid interaction between periodic images in the perpendicular direction. The climbing image-nudged elastic-band (CI-NEB) method\cite{henkelman2000climbing} was employed to search the transition states of the oxygen migration on biphenylene monolayer within the periodic $4\times3$ supercell containing 72 carbon atoms and an oxygen atom.

The thermodynamic stability for oxygen adsorption was measured by the adsorption energy $E_{ad}$, which was defined as $E_{ad} = E_{O@bip} - E_{bip} - E_O$, where, $E_{O@bip}$, $E_{bip}$ and $E_O$ were the energies of the biphenylene monolayer with and without the adsorbed oxygen atom and an isolated oxygen atom, respectively. The potential energy surface (PES) was plotted by calculating the geometric relaxation energy of an oxygen atom on the grid surface of biphenylene monolayer. The positions of biphenylene monolayer and the oxygen atom were fixed in the plane, allowing their relaxation in the \textit{z}-direction to reach the adsorption equilibrium.\cite{brocks1991binding,jackle2014microscopic} In order to reduce the calculation cost while retaining the accuracy, we used a $3\times2$ supercell model with 36 carbon atoms and an oxygen atom and calculated the irreducible part of the biphenylene monolayer. The Brillouin zone was sampled with a Gamma-centered $2\times2\times1$ \textit{k}-point grids. A high-density grid of 460 points was used to scan all the possible sites, with a grid point spacing of about 0.1 \AA\ in the \textit{x} and \textit{y} directions.

\begin{figure*}[bp]
	\centering
	\includegraphics[width=0.75\textwidth]{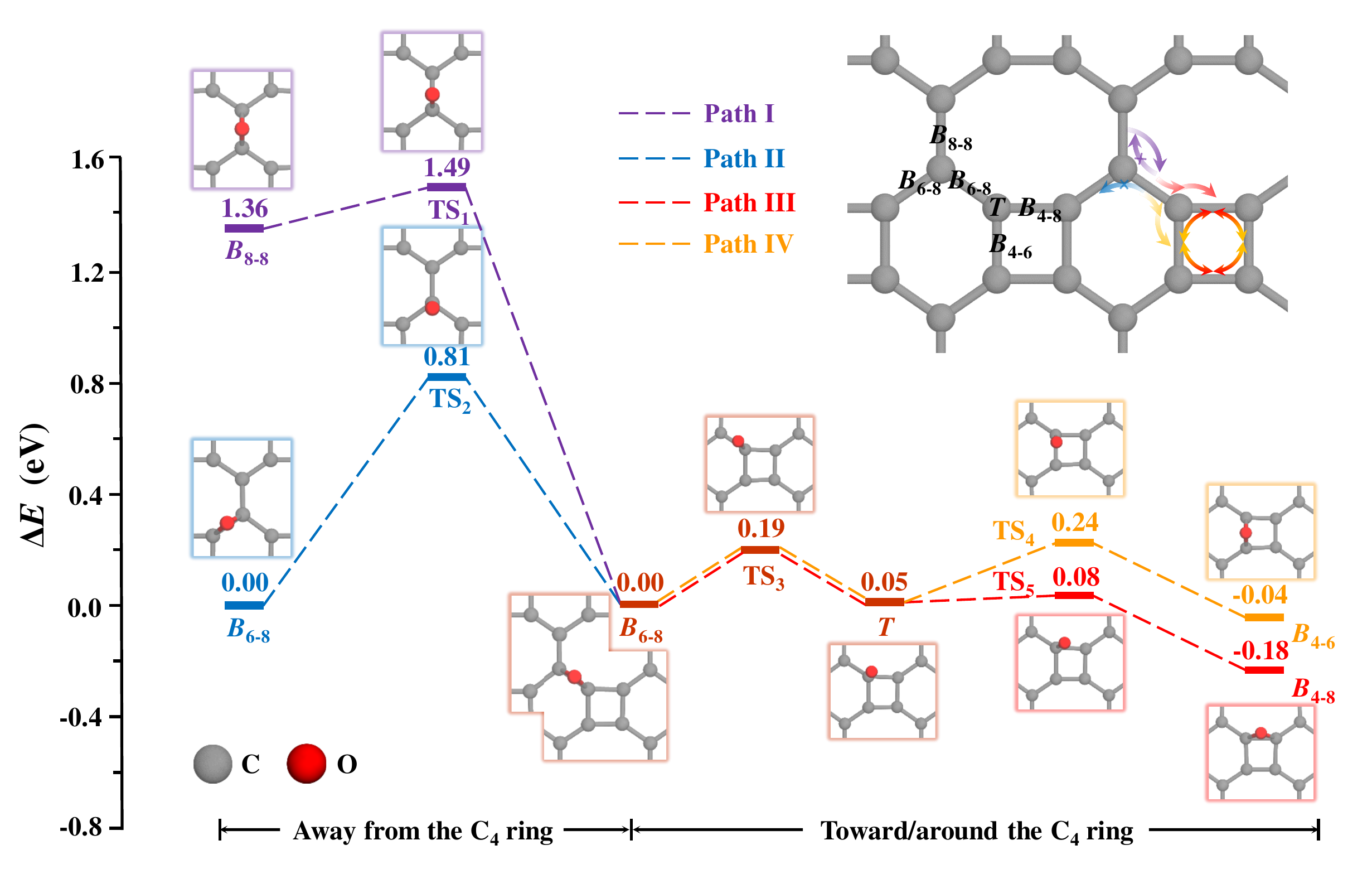}
	\caption{The oxygen migration pathways and state configurations on the biphenylene monolayer. The ${\Delta}E$ is the relative energy with respect to the oxygen chemisorption energy at the $B_{6-8}$ site. Four migration pathways connecting five oxygen chemisorption sites are denoted by Path \Rmnum{1} (purple line), Path \Rmnum{2} (blue line), Path \Rmnum{3} (red line) and Path \Rmnum{4} (orange line), respectively. The inset illustrates the six oxygen chemisorption sites and four oxygen migration pathways. The non-equivalent carbon sites are labeled by the same notations used in Fig. 2.}
	\label{figure3}
\end{figure*}

MLMD simulations were performed under the canonical ensemble (NVT)\cite{martyna1992nose} via the machine-learning-based force field (MLFF) recently implemented in VASP 6.3.\cite{jinnouchi2019fly,jinnouchi2019phase} The MLFF was based on \textit{ab initio} molecular dynamics simulations, and the machine judged whether the current MLFF was suitable for the configuration at the current step. If the error was high, the MLFF calculation was stopped and the \textit{ab initio} calculation was performed to obtain accurate energies, forces and stresses, then the configuration was added to the training set and the MLFF was updated. This approach dramatically increased the size and time scale of molecular dynamics simulations while retaining the first-principles accuracy\cite{behler2016perspective,bartok2017machine} and has been applied to the study of phase transition,\cite{liu2022phase} heat transport\cite{verdi2021thermal} and hydration free energies.\cite{jinnouchi2021first} To build a machine-learned interatomic potential, we set up ten different periodic structures with 72 carbon atoms and 4 randomly distributed oxygen atoms at 300 K. For each configuration, the training process was performed for 20 ps with a time step of 0.5 fs to obtain the local reference configurations as many as possible. Then, 762 structures were automatically selected to construct the MLFF with the root-mean-square errors in the energy of 2.5 meV per atom. The temperature of MLMD simulations for dynamic oxygen migration on biphenylene monolayer was maintained at 300 K by the $Nos\acute{e}$-$Hoover$ thermostats with a time step of 0.5 fs. The total simulation time was 50 ps and data were collected after 5 ps when the system has reached the equilibrium. The VASPKIT code\cite{wang2021vaspkit} and VMD\cite{humphrey1996vmd} are used for data processing and graphics production.

\section{Results and discussions}

We first explore the possible chemisorption configurations of an oxygen atom on the biphenylene monolayer and their thermodynamic stability. As shown in Fig. \ref{figure2}, five different chemisorption configurations are found, including C-O-C on the shared C-C bond between adjacent $C_8$-$C_8$ ($B_{8-8}$ site), $C_6$-$C_8$ ($B_{6-8}$ site), $C_4$-$C_8$ ($B_{4-8}$ site) and $C_4$-$C_6$ ($B_{4-6}$ site) rings, and a dangling C-O bond on the shared carbon atom of $C_4$-$C_6$-$C_8$ rings ($T$ site). These configurations correspond to five different non-equivalent carbon sites. Another site, the shared carbon atom of $C_6$-$C_8$-$C_8$ ring, is unstable for oxygen chemisorption and is not shown. In these configurations, the carbon atom is attracted out of the monolayer by the oxygen atom, resulting in a local buckling induced by the bonding change from $sp^2$ to $sp^3$-like hybridization. The $E_{ad}$ for epoxy near/around the $C_4$ ring shows the lower value from -3.53 eV to -3.30 eV while on the $B_{8-8}$ site has the highest value of -1.99 eV. This indicates that the oxygen atom is more preferably chemisorbed near/around the $C_4$ ring. It should be noted that the dangling C-O bond can be stabilized on the biphenylene monolayer, while is unstable on the pristine graphene.\cite{zhang2016substrate,nguyen2013ab} The larger adsorption energy of oxygen sites near/around the $C_4$ ring, compared to that of -2.51 eV on the pristine graphene, and the existence of the stable dangling C-O bond, all suggest that the hybrid carbon rings structure of biphenylene monolayer has a significant impact on the oxygen chemisorption.

\begin{figure*}[bp]
%	\begin{wrapfigure}{l}{0.60\linewidth}
		\centering
		\includegraphics[width=0.60\linewidth]{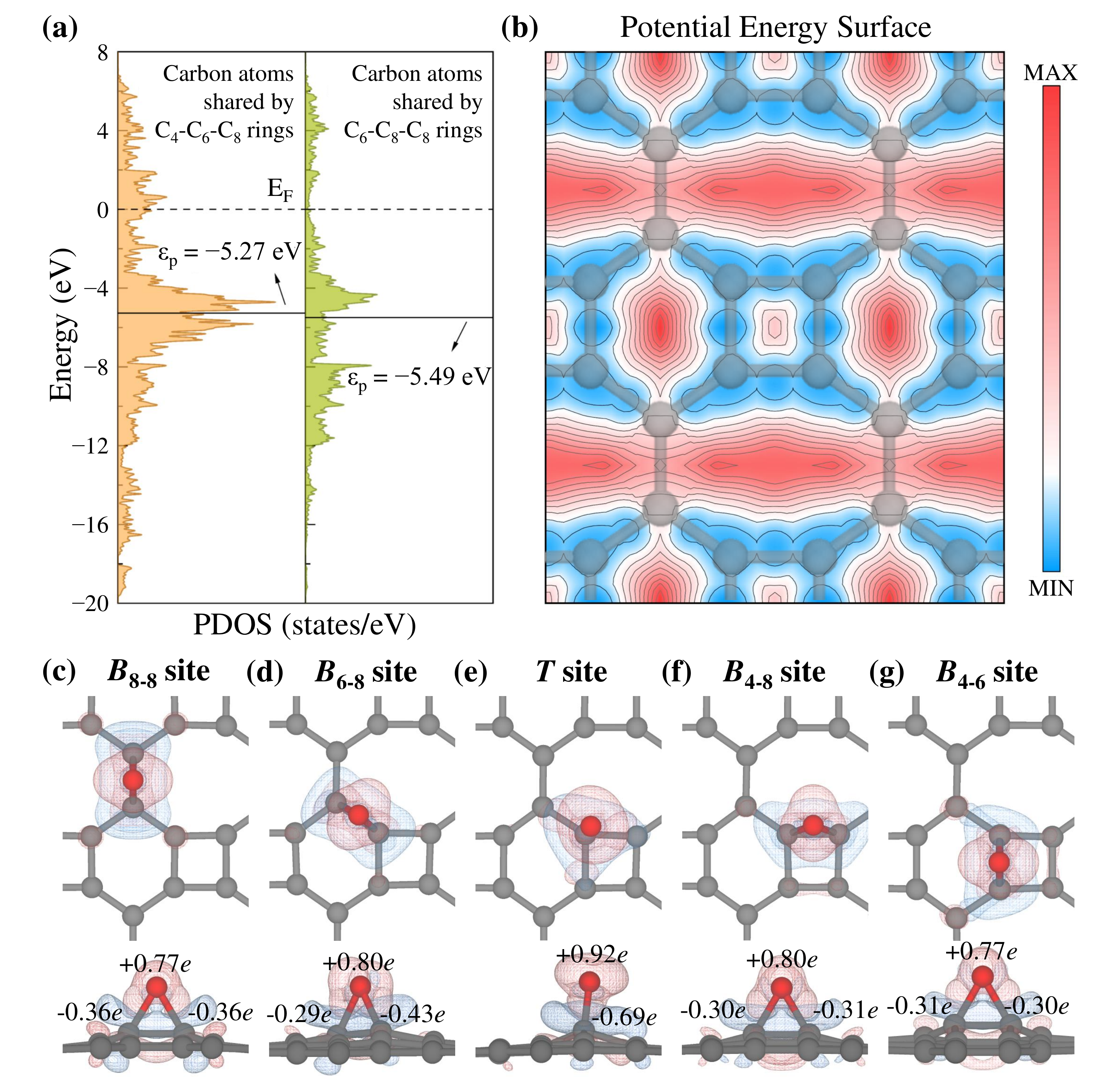}
		\caption{(a) Projected density of states (PDOS) of two types of carbon sites, carbon atoms shared by $C_4$-$C_6$-$C_8$ rings and $C_6$-$C_8$-$C_8$ rings, on the pristine biphenylene monolayer. $\epsilon_p$ and $E_F$ denote the \textit{p}-band center and Fermi level, respectively. (b) The contour map of a potential energy surface (PES) of an oxygen atom adsorbed on the biphenylene monolayer. The dark red and dark blue areas indicate the higher and lower values of adsorption energies, respectively. The contour spacing is 0.25 eV. The charge density difference and Bader charge of the oxygen atom chemisorbed at the (c) $B_{8-8}$, (d) $B_{6-8}$, (e) T, (f) $B_{4-8}$ and (g) $B_{4-6}$ sites. Blue and red colors represent the charge depletion and accumulation regions with the isosurface of 0.005 $e/bohr^3$, respectively. The number indicates the value of Bader charge, with positive and negative values denoting the gain and loss of charge with respect to the isolated atom.}
		\label{figure4}
%	\end{wrapfigure}
\end{figure*}

The different chemisorption configurations generated by the hybrid carbon rings structure of biphenylene monolayer lead to various oxygen migration pathways, and it is found that the oxygen tends to spontaneously migrate toward and around the $C_4$ ring. As shown in Fig. \ref{figure3}, Path \Rmnum{1} (between $B_{8-8}$ and $B_{6-8}$ sites), Path \Rmnum{2} (between $B_{6-8}$ and $B_{6-8}$ sites), Path \Rmnum{3} (between $B_{6-8}$ and $B_{4-8}$ sites) and Path \Rmnum{4} (between $B_{6-8}$ and $B_{4-6}$ sites) are four elementary oxygen migration pathways connecting five different oxygen chemisorption sites. Paths \Rmnum{1} and \Rmnum{2} involve a C-O bond breaking in epoxy followed by the formation of an unstable dangling C-O bond in the transition state and the C-O bond formation. Both of these two paths show relatively high barriers of 1.49 eV and 0.81 eV for the C-O bond breaking (from $B_{6-8}$ site to $TS_1$ and $TS_2$), compared to the order of thermal fluctuations at room temperature. This indicates that the oxygen atom is difficult to depart away from the $C_4$ ring along Paths \Rmnum{1} and \Rmnum{2}. However, the oxygen atom can spontaneously migrate from $B_{8-8}$ site toward the $C_4$ ring due to a relatively low barrier of 0.13 eV along the Path \Rmnum{1}. Paths \Rmnum{3} and \Rmnum{4} are two pathways around the $C_4$ ring, and the oxygen migration from $B_{6-8}$ site to $B_{4-6}$ site or $B_{4-8}$ site is assisted with a stable dangling C-O bond at $T$ site. The low barriers for C-O bond breaking (from $B_{6-8}$ site to $TS_3$, 0.19 eV), C-O bond reforming (from $T$ site to $TS_4$ and $TS_5$, 0.19 eV and 0.03 eV, respectively) and their reverse processes (0.14 eV from $T$ site to $TS_3$, 0.28 eV from $B_{4-6}$ site to $TS_4$, 0.26 eV from $B_{4-8}$ site to $TS_5$) suggest the spontaneous oxygen migration toward/around the $C_4$ ring.

\begin{figure*}[t]
	\centering
	\includegraphics[width=0.73\textwidth]{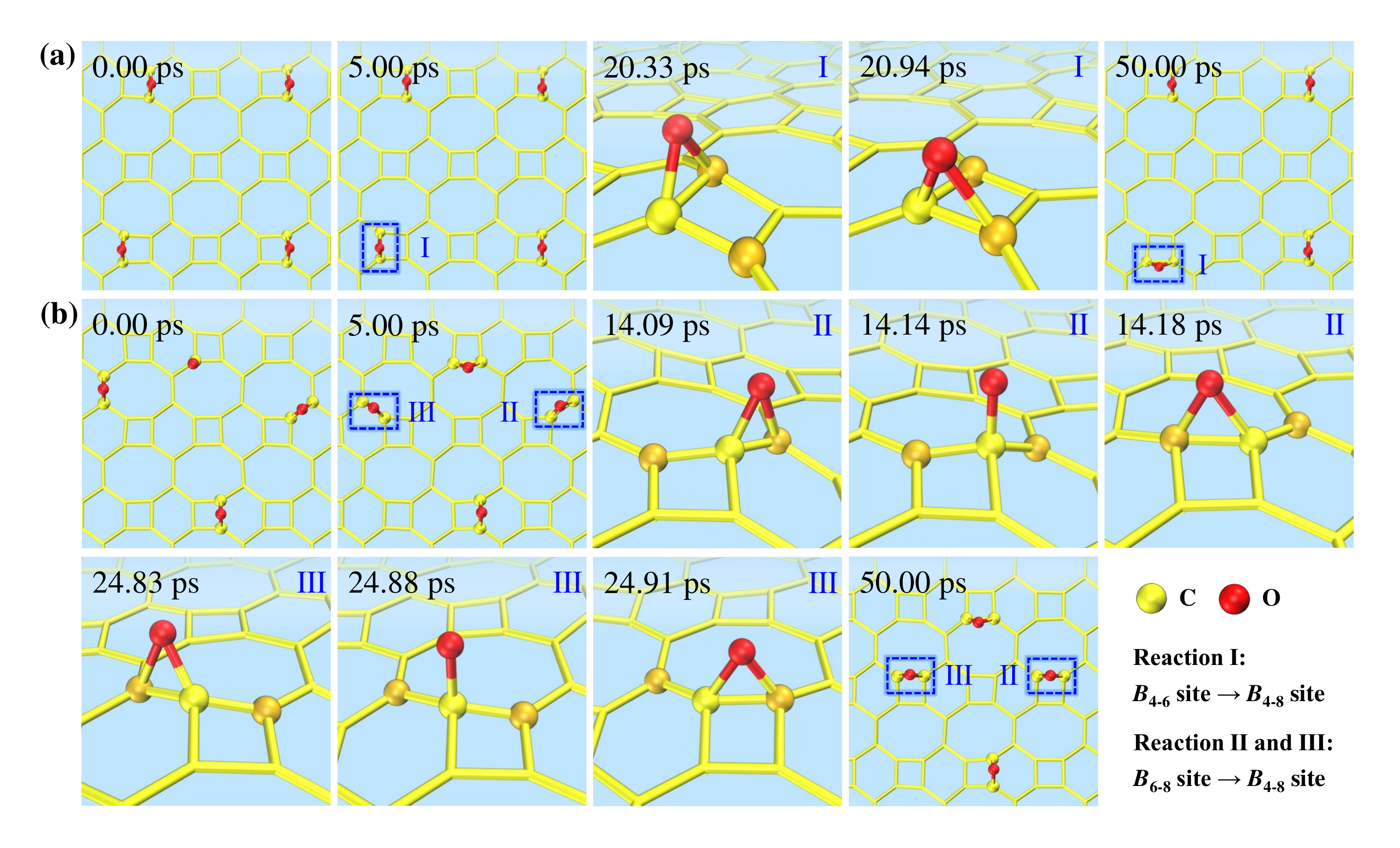}
	\caption{(a, b) Two MLMD trajectories of oxygen migration on the biphenylene monolayer. Three reactions are denoted by Reactions \Rmnum{1} (from $B_{4-6}$ site to $B_{4-8}$ site), \Rmnum{2} and \Rmnum{3} (from $B_{6-8}$ site to $B_{4-8}$ site). The numbers in the upper left of snapshots indicate the time. The periodic display is used in snapshots to get a better view.}
	\label{figure5}
\end{figure*}

Now we wish to analyze the physical mechanism underlying the locally spontaneous dynamic oxygen migration toward/around the $C_4$ ring. Fig. \ref{figure4}(a) presents the projected density of states (PDOS) of two types of carbon sites on the pristine biphenylene monolayer. The \textit{p}-band center ($\epsilon_p$) of the carbon atom shared by $C_4$-$C_6$-$C_8$ rings ($T$ site, -5.27 eV) is closer to Fermi level ($E_F$) than that of the carbon atom shared by $C_6$-$C_8$-$C_8$ rings (-5.49 eV), implying a higher chemical activity of carbon atoms on the $C_4$ ring in the oxygen chemisorption.\cite{liu2021two} The PES, as shown in Fig. \ref{figure4}(b), gives a landscape of interaction between the oxygen atom and the biphenylene monolayer. It can be seen clearly that the existence of five minima, four of which (dark blue) are located at the $B_{6-8}$, $B_{4-8}$, $B_{4-6}$ and $T$ sites near/around the $C_4$ ring. The remanent one is at the $B_{8-8}$ site and the light red color indicates its relatively higher energy with respect to the sites near/around the $C_4$ ring. This is consistent with the thermodynamic stability of oxygen chemisorption sites shown in Fig. \ref{figure2}. Furthermore, we can find that the central areas of carbon rings usually correspond to the maxima and the possible oxygen migration pathways are along the directions of C-C bonds, including pathways between the $B_{8-8}$ and $B_{6-8}$ sites (Path \Rmnum{1}), between two $B_{6-8}$ sites (Path \Rmnum{2}) and around the $C_4$ rings (Paths \Rmnum{3} and \Rmnum{4}). It is obvious that the migration from the $B_{6-8}$ site to $B_{8-8}$ site and another $B_{6-8}$ site is difficult since a saddle point with the relatively high energy is located at the carbon atom shared by $C_6$-$C_8$-$C_8$ rings. This can also be seen from the asymmetric charge density difference and the unequal Bader charges\cite{bader1985atoms} of carbon atoms, -0.29\textit{e} and -0.43\textit{e}, and in the epoxy in Fig. \ref{figure4}(d). The fewer charge of -0.29\textit{e} means the weaker C-O bond strength, and it is much easier for migration toward the carbon atom on the $C_4$ ring ($T$ site) than the carbon atom shared by $C_6$-$C_8$-$C_8$ rings. On the contrary, the spacing between the saddle points (shown by white regions separating four minima) and four minima around the $C_4$ ring is less than a contour spacing, giving an estimated barrier of less than 0.25 eV and indicating the locally spontaneous dynamic oxygen migration around the $C_4$ ring on the biphenylene monolayer.

The locally spontaneous dynamic oxygen migration on the biphenylene monolayer can be further validated by MLMD simulations. Fig. \ref{figure5} presents two MLMD trajectories of oxygen migration starting from different initial configurations with randomly distributed oxygen atoms. In the first trajectory, as shown in Fig. \ref{figure5}(a), from time \textit{t} = 20.33 \textit{ps}, an epoxy at the $B_{4-6}$ site starts to migrate around the $C_4$ ring by breaking a C-O bond. A new epoxy at the $B_{4-8}$ site appears at \textit{t} = 20.94 \textit{ps} when another C-O bond is reformed. This process denoted by Reaction \Rmnum{1} clearly shows the spontaneous dynamic oxygen migration around the $C_4$ ring (from $B_{4-6}$ site to $B_{4-8}$ site). The spontaneous oxygen migration from $B_{4-8}$ site to $B_{4-6}$ site around the $C_4$ ring can be observed in Fig. S1 in ESI. In the second trajectory (see Fig. \ref{figure5}(b)), the epoxy migrates toward the $C_4$ ring starting from $B_{6-8}$ site to $B_{4-8}$ site, with the first case beginning from \textit{t} = 14.09 \textit{ps} to 14.14 \textit{ps} (Reaction \Rmnum{2}) and the second one beginning from \textit{t} = 24.83 \textit{ps} to 24.91 \textit{ps} (Reaction \Rmnum{3}). The oxygen migration presented in MLMD simulations agrees with the pathways based on DFT calculations and confirms the locally spontaneous dynamic oxygen migration toward/around the $C_4$ ring of the biphenylene monolayer.

Given the oxygen migration pathways on the biphenylene monolayer, we expect that the hybrid carbon rings structure may be an effective approach to realize the spatial-controlled dynamic behavior. We evaluate the oxygen migration rate $D$ on the basal plane of biphenylene monolayer according to $D = {\nu}d^2exp(-\epsilon/k_BT)$,\cite{meunier2002ab} where $\nu$ is the attempt frequency,\cite{koettgen2017ab} $d$ is an elementary migration length, $\epsilon$ is the energy barrier, $k_B$ is the Boltzmann constant, and $T$ is the absolute temperature (see details in Table S1 in ESI). It is found that the low energy barriers toward/around the $C_4$ ring resulted in the fast migration rate ranging from the order of $10^{-8}$ to $10^{-5}$ $cm^2/s$ at 300 K. Conversely, the oxygen migration rate departing away from the $C_4$ ring is 9--23 orders of magnitude lower than that toward/around the $C_4$ ring. Therefore, the oxygen atom is in fact trapped around the $C_4$ ring, and the long-distance migration is extremely difficult.

\section{Conclusions}
In summary, based on DFT calculations and MLMD simulations, we have demonstrated the locally spontaneous dynamic oxygen migration on the biphenylene monolayer. Various oxygen migration pathways connecting the different oxygen chemisorption configurations are shown, and it is found that the oxygen atom tends to spontaneously migrate toward/around the $C_4$ ring rather than departs the $C_4$ ring. This means the locally spontaneous dynamic oxygen migration on the biphenylene monolayer, which is attributed to the low barrier of about 0.3 eV for the former process and relatively high barrier of about 1.5 eV for the latter one, originating from the enhanced activity of C-O bond near/around the $C_4$ ring caused by the hybrid carbon rings structure. Furthermore, the MLMD simulations are consistent with the oxygen migration pathways calculated from DFT calculations, and confirm the locally spontaneous dynamic oxygen migration toward/around the $C_4$ ring on the biphenylene monolayer. 

The locally spontaneous dynamic behavior of oxygen atom on the biphenylene monolayer lays the foundation for future studies of oxygen-related reactions and global dynamic properties of biphenylene monolayer. In fact, it has been reported that the biphenylene monolayer shows great potential as metal-free catalysts for oxygen reduction and oxidation reactions.\cite{su2022theoretical,liu2021two} Therefore, the localized dynamic behavior of oxygen atom near/around the $C_4$ ring provides possibility of spatial-controlled catalytic reactions. Moreover, our work emphasizes the role of hybrid carbon rings on the interfacial oxygen activity, and gives the framework to understand the dynamic behavior of oxygen on other 2D carbon allotropes, such as T-graphene,\cite{liu2012structural} penta-graphene\cite{zhang2015penta} and net-graphene.\cite{liu2017graphene}

\section*{Author Contributions}
Y. Tu and Z. Yan conceived, designed and guided the research; B. Situ, Z. Yan and R. Huo performed the simulations; B. Situ, Z. Yan, K. Wang, L. Chen, Z. Zhang, L. Zhao and Y. Tu analyzed the data; B. Situ, Z. Yan and L. Zhao wrote the paper; All the authors participated in discussions of the research.

\section*{Conflicts of interest}
There are no conflicts to declare.

\section*{Acknowledgements}
We are thankful for the helpful suggestions given by Zhijing Huang (Yangzhou University), Wenjie Yang (University of Science and Technology of China) and Hao Yang (Yangzhou University). Zihan Yan thanks Lin Zhao (NanJing XiaoZhuang University) for drawing the TOC. This work was funded by the National Natural Science Foundation of China (Nos. 12075201, 11675138, 11605151), the Natural Science Foundation of Jiangsu Province (No. BK20201428) and Jiangsu Students’ Innovation and Entrepreneurship Training Program (No. 202211117016Z).

%%%END OF MAIN TEXT%%%


\balance

\renewcommand\refname{References}

%%%REFERENCES%%%
%\bibliography{rsc} %You need to replace "rsc" on this line with the name of your .bib file
%\bibliographystyle{rsc} %the RSC's .bst file

\providecommand*{\mcitethebibliography}{\thebibliography}
\csname @ifundefined\endcsname{endmcitethebibliography}
{\let\endmcitethebibliography\endthebibliography}{}

\end{document}